%% file: Baczko_AGN_jets.tex
\documentclass[a4paper,11pt]{article}
\usepackage{aaskaiid}
\usepackage{wrapfig}
\usepackage{orcidlink}
\title{AGN Jets from Formation to Dissipation}
\ShortTitle{AGN Jets from Formation to Dissipation}

\ShortName{Baczko, A.-K. et al.} 
\author[1,2]{A.-K. Baczko \orcidlink{0000-0003-3090-3975}}
\author[3,4]{Eleni Vardoulaki\orcidlink{0000-0002-4437-1773}}
\author[5,6,7]{Etienne Bonnassieux}
\author[8,9]{Manel Perucho}
\author[7]{Marisa Brienza \orcidlink{0000-0003-4120-9970}}
\author[10]{Marcus Br\"uggen}
\author[11,12]{Emmanuel K. Bempong-Manful \orcidlink{0000-0002-1727-1224}}
\author[11]{Robert Beswick}
\author[13,2,14]{Florian Eppel \orcidlink{0000-0001-7112-9942}}
\author[6]{Damien Gratadour}
\author[13]{Jonas Heßdörfer\orcidlink{0009-0009-7841-1065}}
\author[8]{Kiara Hervella-Seoane}
\author[13]{Matthias Kadler \orcidlink{0000-0001-5606-6154}}
\author[15]{Jae-Young Kim \orcidlink{0000-0001-8229-7183}}
\author[16,17,18]{Aretaios Lalakos}
\author[19,20]{Leah K. Morabito \orcidlink{0000-0003-0487-6651}}
\author[21,2]{Dhanya G. Nair  \orcidlink{0000-0001-5357-7805}}
\author[22,23]{Felix P\"otzl} 
\author[11]{Jack Radcliffe}
\author[13]{Luca Ricci \orcidlink{0000-0002-4175-3194}}
\author[2]{Eduardo Ros}
\author[24]{Jan R\"oder \orcidlink{0000-0002-2426-927X}}
\author[13]{Florian R\"osch \orcidlink{0009-0000-4620-2458}}
\author[13]{Ainara Saiz-P\'erez}
\author[13]{Hrishikesh Shetgaonkar}
\author[25,26]{Cyril Tasse} 
\author[27,28]{Francesco Ubertosi\orcidlink{0000-0001-5338-4472}}
\author[1]{Jun Yang}
\author[2]{J. Anton Zensus}

\affiliation[1]{Department of Physics and Astronomy, Chalmers University of Technology, SE-439 92 Onsala, Sweden}
\emailAdd{anne-kathrin.baczko@chalmers.se}
\affiliation[2]{Max-Planck-Institut f{\"u}r Radioastronomie, Auf dem H{\"u}gel 69, D-53121 Bonn, Germany}
\affiliation[3]{National Observatory Athens, Hill of the Nymphs, Athens, Greece}
\emailAdd{elenivard@gmail.com}
\affiliation[4]{Th{\"u}ringer Landessternwarte, Sternwarte 5, 07778 Tautenburg, Germany}
\affiliation[5]{Laboratoire d’Astrophysique de Bordeaux, Univ. Bordeaux, CNRS, B18N, all\'ee Geoffroy Saint-Hilaire, 33615 Pessac, France
}
\emailAdd{etienne.bonnassieux@u-bordeaux.fr}
\affiliation[6]{LUX, Observatoire de Paris, PSL, CNRS, SU/UPMC, UPD, 5 place Jules Janssen, 92195 Meudon, France}
\affiliation[7]{INAF - Istituto di Radioastronomia, via Gobetti 101, 40129 Bologna, Italy}
\affiliation[8]{Departament d’Astronomia i Astrofísica, Universitat de València, Av/ Vicent Andr\'es Estell\'es, 18, 46100, Burjassot, Val\`encia, Spain}
\emailAdd{manel.perucho@valencia.edu}
\affiliation[9]{Observatori Astronòmic, Universitat de València, C/ Catedràtic José Beltrán 2, 46980, Paterna, Val\`encia, Spain}
\affiliation[10]{Hamburger Sternwarte, Universität Hamburg, Gojenbergsweg 112, 21029 Hamburg, Germany}
\affiliation[11]{Jodrell Bank Centre for Astrophysics, Department of Physics and Astronomy, The University of Manchester, Manchester M13 9PL, UK}
\affiliation[12]{School of Physics, University of Bristol, Tyndall Avenue, Bristol BS8 1TL, UK}
\emailAdd{emmanuel.bempong-manful@manchester.ac.uk}
\affiliation[13]{Julius-Maximilians-Universit{\"a}t W{\"u}rzburg, Fakult{\"a}t für Physik und Astronomie, Institut für Theoretische Physik und Astrophysik, Lehrstuhl für Astronomie, Emil-Fischer-Str. 31, D-97074 W{\"u}rzburg, Germany}
\affiliation[14]{Joint Institute for VLBI ERIC, Oude Hoogeveensedijk~4, 7991 PD Dwingeloo, The Netherlands}
\affiliation[15]{Department of Physics, Ulsan National Institute of Science and Technology (UNIST), 50 UNIST-gil, Eonyang-eup, Ulju-gun, Ulsan 44919, Republic of Korea}
\affiliation[16]{TAPIR, Mailcode 350-17, California Institute of Technology, Pasadena, CA 91125, USA}
\affiliation[17]{Walter Burke Institute for Theoretical Physics, California Institute of Technology, Pasadena, CA 91125, USA}
\affiliation[18]{Canadian Institute for Theoretical Astrophysics, 60 St. George Street, Toronto, ON M5S 3H8, Canada}
\affiliation[19]{Centre for Extragalactic Astronomy, Department of Physics, Durham University, South Road, Durham DH1 3LE, United Kingdom}
\affiliation[20]{Institute for Computational Cosmology, Department of Physics, Durham University, South Road, Durham DH1 3LE, United Kingdom}
\affiliation[21]{Astronomy Department, Universidad de Concepci\'{o}n, Casilla-160c, Concepci\'{o}n, Chile}
\affiliation[22]{Institute of Astrophysics, Foundation for Research and Technology – Hellas, N. Plastira 100, Voutes, GR-70013, Heraklion, Greece}
\affiliation[23]{University of Crete, Department of Physics \& Institute of Theoretical \& Computational Physics, 70013, Heraklion, Greece}
\affiliation[24]{Instituto de Astrofísica de Andalucía-CSIC, Glorieta de la Astronomía s/n, 18008 Granada, Spain }
\affiliation[25]{GEPI, Observatoire de Paris, Universié PSL, CNRS, 5 Place Jules Janssen, 92190 Meudon, France}
\affiliation[26]{Department of Physics \& Electronics, Rhodes University, PO Box 94, Grahamstown 6140, South Africa }
\affiliation[27]{Dipartimento di Fisica e Astronomia, Università di Bologna, via Gobetti 93/2, I-40129 Bologna, Italy}
\affiliation[28]{Istituto Nazionale di Astrofisica - Istituto di Radioastronomia (IRA), via Gobetti101, I-40129 Bologna, Italy}

\abstract{Active galactic nuclei (AGN) are among the most energetic phenomena in the Universe, capable of launching powerful relativistic jets that extend from sub-parsec to megaparsec scales. These jets play a crucial role in regulating star formation, redistributing energy and matter, and shaping the evolution of galaxies and their environments. Despite decades of study, a comprehensive understanding of how AGN jets form, propagate, and dissipate remains elusive. The aim of this chapter is to highlight how the future capabilities of the the Square Kilometre Array (SKA), as a standalone array as well as in combination with Very Long Baseline Interferometry (VLBI) arrays and multi-wavelength facilities, will transform our capabilities to study the co-evolution of AGN jets and their host galaxies from jet formation to dissipation scales.}

\begin{document}
\include{journal-names}
\maketitle

\section{Introduction}

\begin{figure}[!hb]
\centering
    \includegraphics[width=0.98\textwidth]{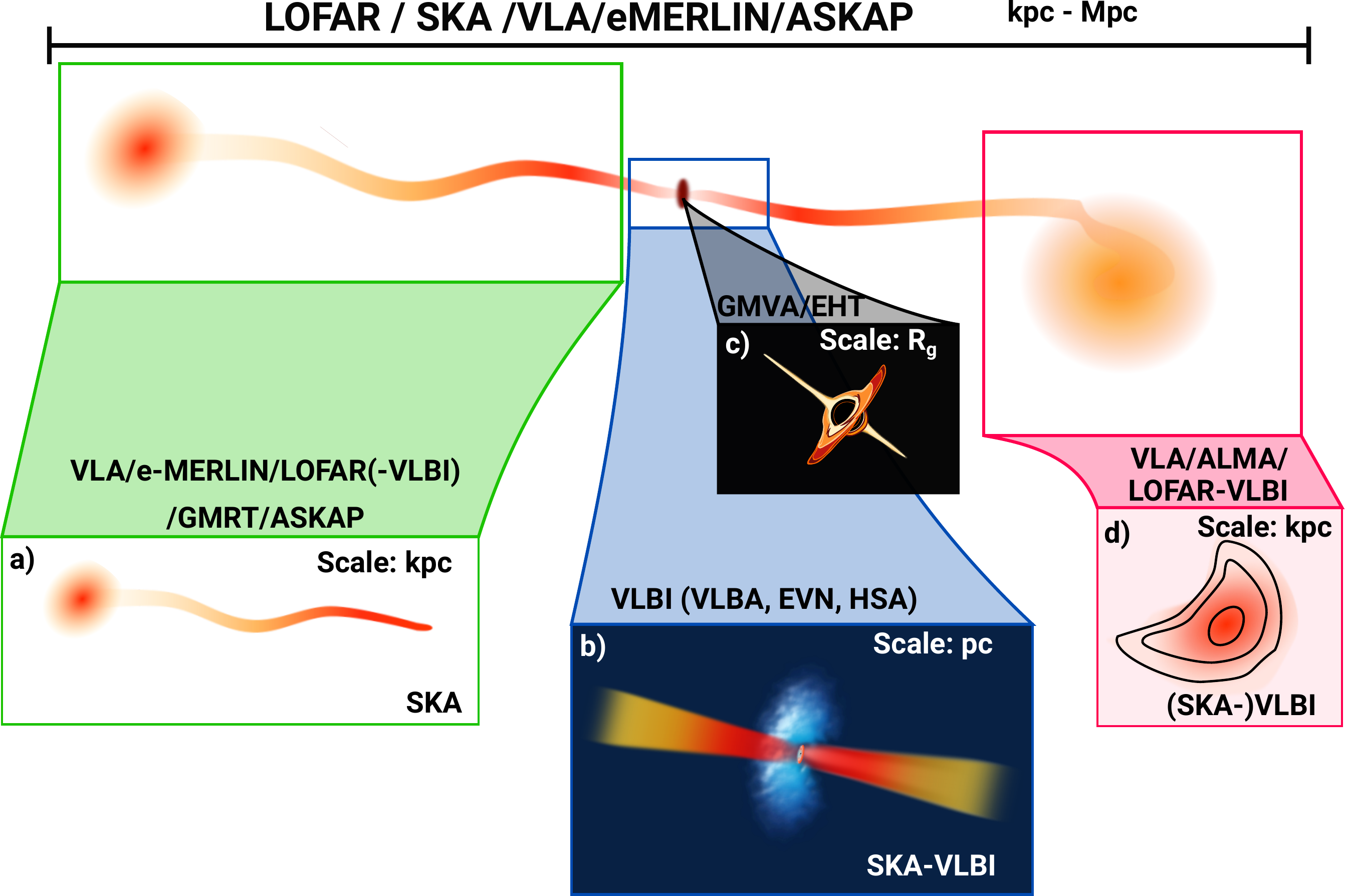}
  \caption{\textit{Sketch of all scales of AGN jets observable at radio wavelengths. a) jet propagation and interaction region; b) jet formation and launching region; c) black hole accretion and jet formation region d) jet dissipation and feedback region}}
  \label{fig:chapter_overview}
\end{figure}

Active Galactic Nuclei (AGN) are among the most energetic phenomena in the Universe. Powered by
accretion onto supermassive black holes (SMBHs), AGN can launch relativistic jets that regulate star formation
and redistribute energy across their host galaxies. The unified model explains the diversity of AGN \citep[e.g.][]{Bla19} via orientation of a central SMBH, an accretion disk, a torus, and bipolar jets. We show in Figure~\ref{fig:chapter_overview} an overview of the different scales relevant to AGN evolution and which radio facilities allow us to probe these scales.

Radio astronomy has been instrumental in probing different aspects of AGN jet physics. Very Long Baseline Interferometry (VLBI) has enabled detailed studies of the innermost jet regions, close to the central SMBH, where jets are launched and collimated \citep{EHT19a, Lu23} (see Fig.~\ref{fig:chapter_overview}c). Large-scale radio surveys with connected element interferometers, like VLA and LOFAR, have revealed the impact of jets on their host galaxies and the surrounding interstellar and circumgalactic media \citep{Shimwell2019} (see Fig.~\ref{fig:chapter_overview}a and d). However, current instruments lack the sensitivity and resolution to simultaneously study the full range of physical scales involved in AGN jet evolution, limiting our ability to physically describe the impact of AGN jet activity on the host galaxy and ultimately on galaxy evolution.

The SKA, with its combination of unprecedented sensitivity and angular resolution, and wide frequency coverage, is poised to transform our understanding of AGN jets. In particular, the synergy between SKA-Mid and SKA-VLBI will enable, for the first time, a continuous view of jet structures from sub-parsec to kiloparsec scales. Early science with SKA-Mid/Low Array Assembly~* (AA*) will already allow for detailed spectral and morphological studies of AGN jets complementing past studies performed by existing facilities up to declinations of $+30^{\circ}$, where instruments such as the International LOFAR Telescope and e-MERLIN currently provide observational coverage.

Although AGN spend most of their time in a low luminosity state, current AGN models inadequately represent the faint AGN population of low luminosity AGN (LLAGN), with a bolometric luminosity $\leq 10^{42}\mathrm{erg\,s^{-1}}$, which lack typical torus signatures and show broad-band emission dominated by radiatively inefficient accretion flows (RIAF) and jet emission \citep{Fer23,Jin25}. LLAGN play a crucial role since they are not only underrepresented in existing high-resolution surveys due to their low surface brightness, but they may be the missing link in our understanding of the processes regulating star formation, driving feedback processes, and accelerating cosmic rays \citep[CRs,][]{Rus23} -- key in understanding galaxy formation. The future capabilities with SKA(-VLBI) provide the  resolution and sensitivity to study these sources en masse across all scales.

Finally, SKA-VLBI will open up the southern sky for high-sensitivity milliarcsecond-scale observations, revealing a previously inaccessible population of faint AGN and their jets. While the increased sensitivity of the SKA will allow for the detection of extended, low-brightness regions in the known AGN population, the addition of other arrays to form SKA-VLBI will provide the transformational increase in baseline sensitivity which will allow it to probe, for the first time, the as-yet-undetected AGN population at VLBI scales. All of this will allow us, for the first time, to survey both the faint and the all-sky population of AGN jets to create a complete sample of AGN jets over a wide range of jet power, studied from formation to dissipation. 

This chapter outlines how the SKA, both as a standalone instrument and as part of global VLBI networks, will enable a comprehensive study of AGN jets from formation to dissipation.

\section{Existing radio facilities}

\subsection{The International LOFAR Telescope (ILT)}

The LOw Frequency ARray \citep{2013A&A...556A...2V} is an SKA pathfinder instrument operating in two bands, at 30-80\,MHz and at 124-168\,MHz. It has demonstrated the transformational science potential of deep all-sky surveys, as its combination of a high sensitivity (reaching a median sensitivity of $83\mu$Jy/beam), good angular resolution ($6''$) and large field of view ($\sim5^\circ$) has allowed it to map the entire Northern sky over a period of a decade, providing an invaluable radio counterpart to existing multiwavelength surveys of the Northern sky. In its pathfinder capacity, it has demonstrated the relevance and feasibility of a similar survey conducted using SKA-Low. The full International LOFAR Telescope (ILT) array currently includes baselines up to $\sim2,000\,$km which makes it capable of providing maps with $>6$ deg$^2$ field of view (fov) for an individual field, with sub-arcsecond resolution at 144\,MHz \citep{Sweijen2022,dejong2024,Morabito2025}, which will be extended to 66 MHz thanks to its upcoming LOFAR2.0 upgrade. Thus the ILT provides an exceptional pathfinder for multi-resolution studies of AGN, from large-scale diffuse emission down to sub-galactic and even nuclear scales. In terms of SKA-Low, sub-arcsecond imaging with the ILT has been demonstrated at low declinations \citep[e.g.,][]{Harwood2022}. The ILT will reach the deepest sensitivities possible in the equatorial regions of the sky in combination with SKA-Low. Another exciting opportunity is the Low-frequency Australian Megametre-Baseline Demonstrator Array (LAMBDA), which will provide LOFAR-like capability in the South when coupled with SKA-Low. For more information about the possibilities of SKA-Low within global VLBI arrays we forward the reader to the Chapter on this topic \citep{Timmerman01.2026.SKA}.

\subsection{Multi-scale imaging with existing VLBI arrays}

\begin{figure}
    \centering
    \includegraphics[width=\linewidth]{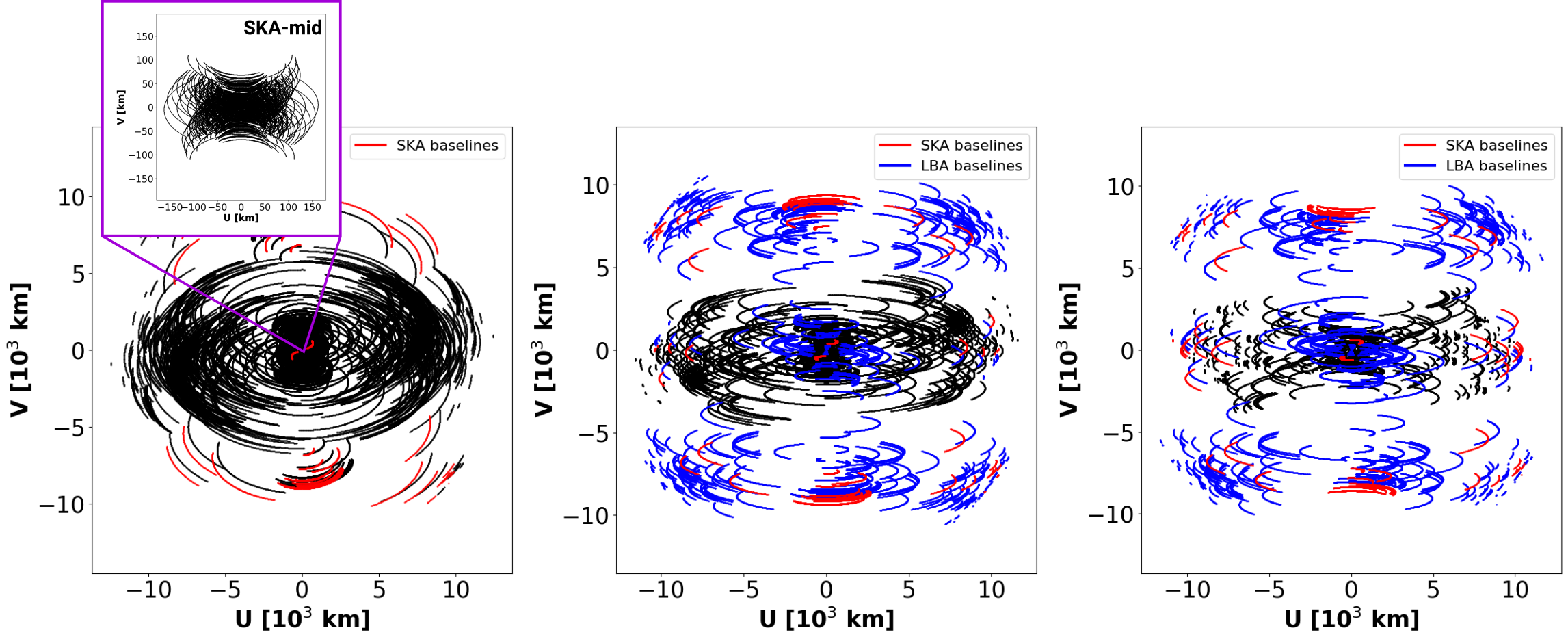}
    \caption{$(uv)$-plot of a global VLBI array consisting of EVN+eMERLIN+VLBA including the SKA (red) for a source at declination of $+30\,$degree (left) and zoom-into the SKA-only coverage, $+10\,$degree (middle), and $-10\,$degree (right). The middle and right plots highlight LBA telescopes as well (blue).}
    \label{fig:uv-plt}
\end{figure}

Current instruments and standard data reduction techniques for VLBI data limit our capability to obtain radio images at intermediate angular resolutions in between parsec and sub-kiloparsec scales. Traditional VLBI arrays, such as the VLBA or the standard EVN, suffer from sparse uv‑coverage and a lack of short baselines, restricting the field of view to only the inner few tens to hundreds of mas. In contrast, connected element interferometers like the JVLA and ALMA provide dense uv‑coverage but lack sufficiently long baselines, yielding large fields of view with inadequate resolution to probe mas scales. Accessing both mas and as structures therefore typically requires combining multiple observations obtained with different instruments and at different frequencies. With SKA‑VLBI, however, the standard SKA‑only data products will be available alongside the VLBI data. This, together with the wide bandwidth will enable simultaneous coverage of both angular‑scale regimes within a single observation. Global observations combining northern VLBI arrays, such as EVN and VLBA, with the southern LBA will benefit largely from the improved sensitivity on long baselines to SKA, reducing the risk of creating disconnected subarrays and as such significantly improving the $(uv)$-coverage. A comparison of current VLBI arrays with and without SKA is shown in figure \ref{fig:uv-plt}.

Studies of the Northern SPARCS (SLOAN) reference field at 1.6\,GHz with the European VLBI Network (EVN) and e-MERLIN telescopes highlight the importance of these intermediate resolutions on which strong interaction between jets and the ambient medium can decelerate the jet plasma to sub-relativistic speeds \citep{Njeri2023}. This not only results in smaller jet structures, but provides an important source of energy transfer from the jets to the inter-stellar medium. The eMERLIN legacy survey of nearby galaxies (LeMMINGs) provides valuable observation and data reduction methods to design a survey of the lower luminosity radio galaxy population on the example of e-MERLIN \citep{Baldi2018}. Studies like these showcase the possibilities of increased scientific output from novel imaging methods using the information from the whole interferometric data in order to recover emission information both from short and long baselines, enabling us to map multiple scales of AGN jets from the same data. The EVN in combination with the e-MERLIN telescopes provide an ideal testbed for developing and improving data reduction methods to allow for multi-scale imaging at GHz frequencies with SKA-Mid. In particular, the significant improvement of bandwidth will allow for multi-frequency synthesis. In combination with improved imaging methods, we will obtain much more information, both spatial and spectral, from the same observation. Existing VLBI stations are actively developing broad band capabilities that are expected to get available within the next years, for example a prototype for the DBBC4 backend providing 28\,GHz bandwidth per IF is expected to be available before the end of 2026 \citep{dbbc4}. Other progress has been made, e.g. with the BRAND receiver, that is under commissioning at the Effelsberg telescopes, followed by the CX receiver at the Yebes telescope, only highlighting some of the upgrades currently undergoing within the EVN. These are great prospects for the usability of the full bandwidths of SKA in future VLBI observations.

During the last years significant progress had been made in developing advanced data reduction and imaging methods for VLBI data employing modern computational and numerical methods. For imaging this includes, e.g., the resolve software which employs Bayesian self-calibration and imaging \citep{Kim2024} or Dob-CLEAN using multiscale and multidirectional wavelet dictionaries \citep{Mueller2023}. The combination of precursor studies for SKA-VLBI, such as possible with EVN+e-MERLIN+MeerKAT+uGMRT, and the continued development of data reduction methods allowing for multi-scale imaging are the ideal base for using the full potential of SKA-VLBI with multiple sub-arrays and multiple tied-array VLBI beams. The ILT, with its multi-beaming capabilities and wide dynamic range of spatial scales provides the ideal testbed for the SKA-Low.

\section{Multi-Scale Jet Science with the SKA}
\subsection{Jet Formation and Launching}

In recent decades, radio astronomy has been very efficient in producing images of the jet acceleration and collimation region by employing the technique of VLBI, from cm-wavelengths with the EVN and Very Long Baseline Array (VLBA) to mm-wavelengths with the Global mm VLBI Array (GMVA) \citep{Ros2024}. The Event Horizon Telescope (EHT) produced the first images of the direct surroundings of central SMBHs \citep{EHT19a}. VLBI observations enabled us to constrain theoretical models of jet formation and collimation by measuring, e.g., the jet expansion within the inner $\sim 10^6 R_\mathrm{S}$ (Schwarzschild radii, e.g. \cite{Pushkarev2017,Bac22}), or by determining the size of the BH shadow\citep{EHTVI}. Characterizing the physical mechanism behind jet collimation is important for the larger goal of studying the impact of the jet on the galaxy evolution. For jets to have an impact on galaxy evolution through energy transfer from the BH into the inter-galactic medium, they need to stay sufficiently collimated \citep{Bla19}. 

The physical processes behind the launching, acceleration, and collimation of these jets remain one of the major open questions even after decades of research \citep[see][and references therein]{Boccardi2017}. The most likely scenario for jet collimation suggests a combination of magnetic self-collimation and a confinement by the ambient medium, which may include the accretion disk (AD) or the torus as well as the interstellar medium \citep[see, e.g.,][]{Spr97,Bes98,Kom07}. 
This is further highlighted by recent results shown in \citet{Boc21}, in which High Excitation Galaxies (HEG) are proposed to stay collimated up to larger scales than Low Excitation Galaxies (LEG). This is suggested to be a direct consequence of HEGs having stronger accretion disk launched winds, highlighting the relevant role of the accretion disk in the collimation of relativistic jets. This is related to the long lasting question on what is the difference between FR\,I and FR\,II jets, is it related to the launching process or differences in the deceleration?
The combination of high-resolution VLBI observations of the parsec scale jet collimation region with detailed knowledge about the molecular content of the host galaxy is key to understanding the delicate interplay between AGN jet evolution and the host galaxy itself.

AGN are expected to spend most of their time in a low luminosity state \cite[e.g.][]{Ho2008}. Thus, current instruments lack the sensitivity to detect significant population sizes, which is required for any statistical study of the lower luminosity end of AGN. The case of the nearby LLAGN NGC\,1052, with its rare double-sided jet, shows how multi-wavelengths VLBI enable us to study the appearance of jet (a-)symmetry of collimation profiles through numerical modeling \citep{Saiz-Perez2025}. Multi-band VLBI observations of NGC\,1052 reveal cylindrical collimation profiles and time-dependent asymmetries \citep{Bac22,Baczko2024,Ric25}, challenging models based solely on magnetic self-collimation \citep{Roh24}. These findings suggest a combined role of magnetic fields and ambient medium confinement, motivating a broader study LLAGN jet morphologies.

Depending on the source properties, such as black hole mass and jet power, the jet collimates on different length scales, after which the expansion breaks into a freely expanding jet. VLBI studies measuring the jet width have shown that this break takes place at a distance between $10^4$ to $10^6\, R_\mathrm{S}$ downstream the central engine. In order to study jet expansion it is critical to measure the jet width over these distances, ideally up to a few orders of $R_\mathrm{S}$ downstream of the break point, until $10^7$ to $10^9\, R_\mathrm{S}$ to allow for significant parameter extraction. This range of covered scales also allows for comparing the location of the break point with the distance of the sphere of influence from the black hole and the bondi radius, which is relevant to study the underlying physical mechanism behind formation an collimation of jets. Presently, this requires the combination of several, different radio facilities to cover the required large range of angular scales, e.g. EVN/VLBA/HSA plus the VLA. SKA-VLBI will change this, thanks to its improved uv coverage resulting from the wide frequency band in a global array, it will be possible to create multiple images from the same observation, recovering different scales, e.g. through different uv-weighting and uv-cuts. SKA-VLBI observations will simultaneously produce standard SKAO data-products, providing us with additional arcsec scale images. This wil be significant, for example in the case of the giant radio galaxy NGC\,315. The combination of observations with the VLBA and the VLA is needed to study the dependence of the jet collimation from the black hole influence, marked as the Bondi radius. While VLBI probe scales upstream of the Bondi radius and reveals a change from a parabolic to a conical expansion profile independent of the Bondi radius, the VLA data probe the jet structures outside the Bondi radius, revealing a continued conical expansion \citep{Boc2021,Park2021}. This and similar studies highlight two distinct limitations of current instruments: 1) At least two different instruments are needed to cover this whole range required to study the dependence of the jet collimation from the black hole influence and to locate possible breaks in the collimation profile; 2) There is limited accessible data available in-between the VLBI and VLA scales, which makes it challenging to continuously trace the jet evolution from the formation region to outside of the Bondi radius. 

SKA, both low and mid, will transform our capabilities to study the jet propagation from pc to kpc scale. Multiple beams will allow for effective observations of faint AGNs by allowing the simultaneous observation of calibration sources and target. Through subarraying and an improved wide bandwidth SKA-VLBI will allow simultaneous observations of multiple angular resolutions down to sub-milliarcseconds. By integrating SKA-Low and SKA-Mid into global VLBI arrays we will further obtain radio signatures of the parsec to kiloparsec jet scales, combining SKA-VLBI with SKA-only, as will be provided by SKAO for each VLBI observation, over a wide frequency range. This will allow spectral index analysis capable of distinguishing between different electron populations, old and new plasma. This is especially relevant for restarted AGN activity and will allow to link jet activity and propagation to AGN duty cycles and galaxy evolution. With present radio facilities it is not possible to study restarted radio galaxies in much detail due to lack of resolution and sensitivity, SKA-VLBI will finally fill this gap of observational capabilities \citep{Bruni2019}. For more information on the topic of jet duty cycle see \citep{Hardcastle01.2026.SKA}.

Finally, recent discoveries of more than dozens of mJy-level faint but large-scale double-lobed jets in spiral and disk galaxies as well as in narrow-line Seyfert~1 (NLSy1) systems \citep[e.g.,][]{Vardoulaki2008, Jarvela2018, Vardoulaki2021a, Keel22,Wu22,Varglund2022,Umayal25} challenge the long-standing view that powerful radio jets arise almost exclusively from massive ellipticals with hot, geometrically thick accretion flows in their centers. 
Furthermore, high-resolution VLBI studies of individual cases reveal parabolic jet collimation profiles connecting sub-parsec and kiloparsec structures, in particular in the spiral galaxy 0313--192 \citep[e.g.,][]{Lee25}. 
On the theoretical side, global GRMHD simulations indicate that radiatively efficient, thin disks can self-generate large-scale poloidal magnetic flux and enter magnetically arrested (MAD) states capable of launching powerful $\Gamma\sim10$ jets \citep[]{Liska19,Liska20,Yosuke22}. 
However, they often fail to produce outflows that can stay collimated to scales beyond kpc, making the origin of their large-scale jets still a mystery.
Therefore, the SKA and SKA-VLBI will be decisive for finding and characterizing those emerging classes of jet systems across cosmic time and understand the global physical mechanisms behind jet formation and propagation.

\subsection{Jet Propagation and Interaction}

Radio AGN jets are complex and dynamic phenomena whose appearance and evolution are governed by their internal structure, particle composition, magnetic fields, and interaction with the surrounding medium. Recent numerical simulations allow us to start probing the complex relation between accretion and jet formation and propagation up to parsec-scales \citep{Lalakos2022}. On parsec to kiloparsec scales, jets are influenced by a variety of physical processes, including instabilities, turbulence, and magnetic fields, which drive energy dissipation and particle acceleration. 

The evolutionary stages of radio AGN can be identified through a combination of morphological and spectral properties. Characterizing their duty cycle is essential for understanding how they influence the evolution of their host galaxies and of their large-scale environment. Parameters such as flux density, size, and accretion rate distinguish high- and low-excitation AGN in bright samples, yet at lower flux densities (of order 100~$\mu$Jy/beam) this separation becomes less clear, and the boundaries between classes blur. For example, high-accretion, compact sources can exhibit low luminosities, while low-accretion objects may still power extended radio structures. Environment also plays a crucial role: FRI-type sources \citep{Fanaroff1974} tend to occupy cosmic filaments \citep{Vardoulaki2021a}, and both compact and giant radio galaxies (GRGs) are found across a wide range of environments, from isolated regions to dense clusters. The largest GRG known, with a projected size of 7~Mpc \citep{Oei2024}, resides in a low-density cosmic web environment, suggesting reduced resistance to jet expansion and demonstrating that jets can remain stable against magnetohydrodynamical instabilities over cosmological distances, even when the Universe was an order of magnitude denser. Conversely, other GRGs, such as those found in the COSMOS field, inhabit group environments with halo masses of $\sim10^{13.3}~M_{\odot}$ \citep{Vardoulaki2025b}, showing sub-Eddington accretion and lower luminosities. These examples highlight that no single evolutionary scenario fits all, and that both jet power and environment jointly shape radio galaxy evolution.

Large-sky surveys such as LoTSS \citep{Shimwell2019,Shimwell2022} and deep-sky surveys like COSMOS \citep{Scoville2007,Vardoulaki2021a} and the Lockman Hole \citep{Sweijen2022,dejong2024} now enable statistical studies of radio AGN properties, relating jet propagation, size, and morphology to host and environmental parameters. However, these studies are typically limited by monochromatic size measurements, which provide only a lower limit on true de-projected source dimensions and hence on age and energetics. The capabilities of SKA-Mid AA*, covering 0.35-15.4~GHz with sub-arcsecond to milliarcsecond resolution (0.3" at 1.4~GHz and 0.03" at 15~GHz), will allow a much more realistic determination of source morphology \cite[see e.g.][]{Sweijen2025} and physical conditions across vastly larger populations of radio AGN.
Observational studies with SKA-pathfinders highlight the potential of SKA for producing spatially resolved spectral ages mapping for radio galaxies \citep{Ubertosi2024}.

By combining wide frequency coverage with the sensitivity to detect faint diffuse emission (tens of $\mu$Jy~beam$^{-1}$), SKA-Mid AA* will bridge the gap between compact and extended components, offering a complete view of how jets evolve and interact across physical scales. This will allow systematic studies of the transition from compact symmetric objects (CSOs) to large-scale FRI, FRII morphologies \citep{Fanaroff1974}, and reveal how energy, magnetic fields, and particle populations evolve during an AGN’s lifetime.

Recent studies have shown that outflows and jet-driven shocks can perturb the interstellar medium (ISM) even in galaxies hosting low-power jets ($L_{\mathrm{jet}}\lesssim10^{44}$~erg~s$^{-1}$), producing velocity spreads of 800–1000~km~s$^{-1}$ perpendicular to the jet axis \citep{Venturi2021}. Such results support cosmological simulations predicting that low-power jets dominate feedback on sub-kiloparsec scales \citep{mukherjee2025}. There is still a lot of uncertainty in low power radio AGN on whether the radio emission is dominated by winds or jets and current facilities do not allow us to distinguish between the two. SKA-Mid’s high spatial and spectral resolution will make it possible to quantify these effects statistically across large, diverse samples and environments.

\begin{figure}
  \vspace{-0.5cm}
  \begin{center}
    \includegraphics[width=\textwidth]{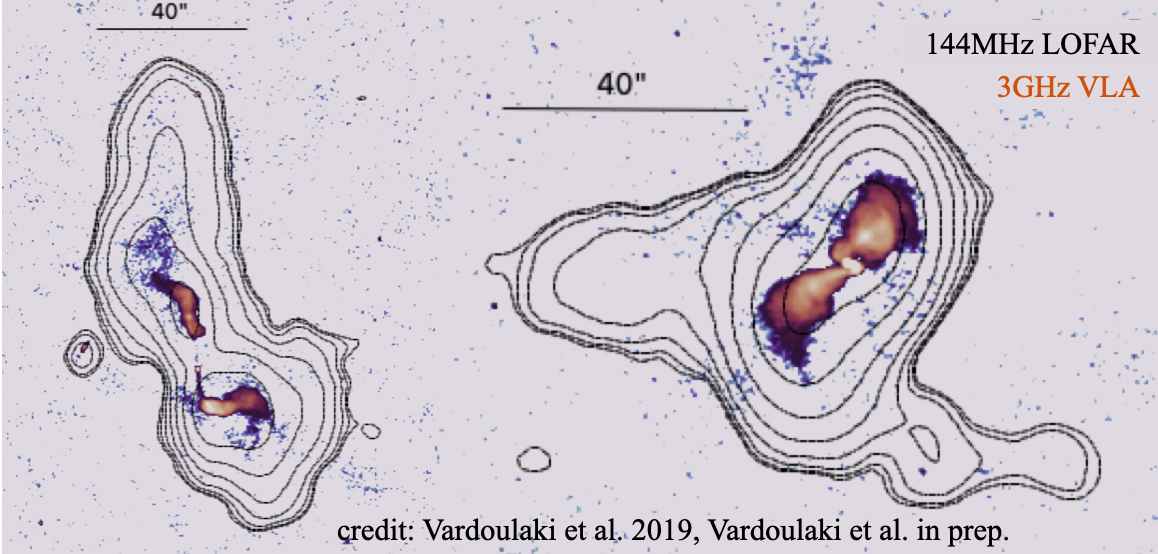}
  \end{center}
  \caption{Examples of two radio AGN at 144 MHz and 3 GHz from the COSMOS field, showing large-scale diffuse emission at MHz frequencies that are missed with GHz observations \citep[][ and Vardoulaki in prep.]{Vardoulaki2021a}}
  \vspace{-0.5cm}
  \label{fig:radioAGN_COSMOS144MHz_3GHz}
\end{figure}

Understanding radio AGN also requires resolving their duty cycle, i.e. the alternation between active, fading, and restarted phases, and linking morphology with spectral ageing. Observations across the full radio spectrum, from tens of MHz to tens of GHz, are essential to capture both fossil and freshly accelerated components. LOFAR-type observations at 60–150~MHz trace remnant plasma and diffuse emission, while GHz observations with MeerKAT \citep[MIGHTEE][]{Hale2025}, ASKAP \citep[e.g. EMU][]{Hopkins2025}, and the VLA probe current jet activity and hotspots (see for example Fig.~\ref{fig:radioAGN_COSMOS144MHz_3GHz}). Existing population studies based on integrated spectral indices reveal a factor of six discrepancy between integrated and resolved radio spectral index measurements \citep{Harwood2017}. While integrated spectra remain valuable due to observational and computational limitations \citep{Vardoulaki2024}, only a few resolved spectral-aging studies exist, typically of bright FRII 3C sources \citep[e.g.][]{Harwood2013,Harwood2017,Bonnassieux2022,Mahatma2023}, as well as of remnants and FRI sources \citep[e.g.][]{Shuleveski2017,Brienza2020,Brienza2022,Brienza2025}. This leaves the faint radio source population largely unexplored, limiting our understanding of their energetics and life cycles. Recent progress combining LOFAR 150~MHz and Apertif 1.4~GHz observations has extended such analyses to restarted and remnant AGN in deep fields like the Lockman Hole \citep{Morganti2021,jurlin2024}, or with LOFAR and GMRT in the XMM-LSS field \citep{Pinjarkar2023}, with similar efforts ongoing in multi-wavelength fields such as COSMOS (Vardoulaki et al., in prep.). SKA-Mid AA* will overcome these limitations by enabling resolved spectral index imaging for millions of sources with sub-arcsecond resolution and $\mu$Jy sensitivity. 

Interpreting these data demands accurate physical modeling. Classical Continuous Injection \citep[CI;][]{jaffe1973} and CI-off models \citep{komissarov1994} describe synchrotron losses under simplified assumptions, but often fail to produce realistic spectra or consistent age estimates \citep{Harwood2017}. Future analyses will benefit from three-dimensional hydrodynamic simulations that explicitly track the evolution and mixing of electron populations within lobes \citep{Yates-Jones2021,Yates-Jones2022}, incorporating environmental asymmetries and adiabatic or radiative losses. These models reproduce observed features such as lobe-length asymmetries and spectral flattening in dense environments, confirming that environment plays a decisive role in radio source evolution. 

Jet–environment interactions span multiple scales. Within the ISM, jets encounter stars and dense gas clouds, generating mixing layers \citep{Perucho2020}; in the circumgalactic medium (CGM), jet feedback regulates star formation by balancing radiative cooling \citep{Prasad2020}; and in the intracluster medium (ICM), jets deposit energy that inflates cavities, drives shock fronts, and generate turbulence \citep[e.g.][]{Ubertosi2023,Prunier2025}. With $\sim$30~mas resolution at 15~GHz, SKA-Mid will directly probe the innermost resolved regions of AGN jets and their coupling to the ISM directly, while sub-arcsecond observations at 1.4~GHz will reveal interactions between jets and circumnuclear gas. These capabilities extend such studies to higher redshifts than previously possible, revealing obscured populations of AGN \citep[e.g.][]{Lambrides2024}, and testing self-regulated feedback mechanisms operating on timescales $\lesssim$100~Myr.

Beyond individual galaxies, SKA-Mid AA* will trace how jets influence their broader environments. Diffuse radio structures such as halos and mini-halos trace re-acceleration of electrons in the intergalactic and intracluster media (IGM and ICM). Measuring their spectral properties will constrain AGN duty cycles and feedback efficiency. For more information, see the Chapter on this topic \citep{Gitti01.2026.SKA}. Observations indicate that radio sources in galaxy groups remain active compared to those in the field \citep{Vardoulaki2025a}, suggesting environmental dependence in AGN evolution. Moreover, high-resolution imaging shows that jet bending can result from ram pressure, but also from interactions with the cooler environment in superclusters with substructures in the warm–hot intergalactic medium \citep[WHIM,][]{Vardoulaki2025b}.  The sensitivity and sky coverage of SKA-Mid will allow such effects to be quantified statistically across cosmic time, while the high resolution will probe the inner regions of bent jets.

Synergistic observations with LOFAR2.0 (20-200~MHz), uGMRT (120~MHz–1.45~GHz), and e-MERLIN (1.25-25~GHz, 0.2"-0.02" resolution), and with EVN/VLBI will enable matched-resolution, multi-frequency studies across the full radio band. Transformative to our understanding of radio AGN at low frequencies have been the past years studies of radio AGN sub-arcsec resolution with LOFAR \citep{Sweijen2022,Morabito2022,Morabito2025}. Together, these facilities will map electron populations continuously from low-frequency fossil plasma to high-frequency emission regions, providing unprecedented insight into jet propagation, feedback, and AGN evolution. Despite the different resolutions, complementary optical, infrared, and X-ray surveys (\textit{Euclid}, \textit{Rubin}, \textit{JWST}, \textit{4MOST}, \textit{Chandra}) will further connect radio AGN to their hosts and environments. Extending the frequency range even higher, synergies with ngVLA (1.2–116~GHz), and the Event Horizon Telescope (EHT) and next-generation EHT (230–345~GHz) will probe jet bases at microarcsecond scales \citep{Baczko2024}, revealing jet structure within the torus and linking small- to large-scale jet propagation. Such observations will clarify why some jets exhibit asymmetries \citep[e.g.][]{Baczko2019}, potentially arising from differences in internal energy, magnetic flux, or kinetic power between the two lobes.

\subsection{Jet Dissipation and Feedback}

VLA, LOFAR and MeerKAT have recently revealed complex structures in jets at very large scales, which are related to the interaction with the IGM \citep[e.g.,][]{2020A&A...636L...1R,2021Galax...9...93V,2025A&A...699A.338H, Brienza2021, Vardoulaki2025a}. Those structures are precisely observed at the low frequency range, and were previously hidden to us. Moreover, the increased sensitivity of those arrays has also opened the path to the discovery of an unexpected number of giant radio galaxies \citep[e.g.,][]{2024A&A...686A..21S,2024Natur.633..537O}. We expect that the increased resolution that SKA will bring, together with its large sensitivity, will represent a major step forward in our understanding of the processes dominating the complex interaction between jets and the multiphase ISM and the IGM \citep[e.g.,][]{2018A&ARv..26....4M,2018MNRAS.479.5544M,2021AN....342.1171P}. It is relevant to understand that these processes can be jet power and medium dependent, so detailed observations of different objects will be necessary precisely to quantify this dependency \citep{2024A&A...684A..45P}. 

This information is crucial to understand the role played by jets in galaxy evolution. Although this is expected to happen via the removal of the ISM gas by the jet-triggered shock and the resulting drop in star formation \citep[][]{2007ARA&A..45..117M,2012ARA&A..50..455F}, it is still under debate whether in some cases, the compression of cold gas in ISM clouds could result in a short burst of enhanced stellar births \citep[e.g.,][]{2012MNRAS.425..438G,2012ApJ...757..136W,2024MNRAS.531.2079M}. In order to study these phenomena in detail and their relative relevance throughout the evolution of our Universe, we need both detailed observations of individual galaxies and large surveys that allow us to perform statistical studies \citep[e.g.,][]{2019A&A...622A..12H,Vardoulaki2021a}. Nevertheless, it seems clear that the interaction of compact jets (GPS, CSS sources with linear sizes of a few kpc) with the multi-phase ISM within the host galaxy triggers a plethora of effects, such as shock ionization or atomic and molecular outflows, depending on jet power and narrow line region cloud properties \citep[e.g.][Bonanad-Hurtado et al., in prep.]{2018A&ARv..26....4M,2024A&A...684A..45P,2024MNRAS.531.2079M,mukherjee2025}. These interactions can allow us to study the connection between the non-thermal and thermal phases in jetted AGN as jets deposit a relevant fraction of their energy flux into the ISM.    

With the aim of understanding the relevance of galactic activity in the evolution of galaxies, it is fundamental to accumulate large amount of data that permit to characterize both jet and host galaxy properties. These can then be used as inputs for Monte Carlo/machine learning simulations of jet evolution which can be compared with the observed samples. In particular, this can shed light on jet power distributions and active galaxy occurrence as a function of redshift \citep[][Beltran-Palau et al., in prep.]{2025arXiv250924939B}. 

\begin{figure}[!hb]
    \centering
    \includegraphics[width=0.44\linewidth]{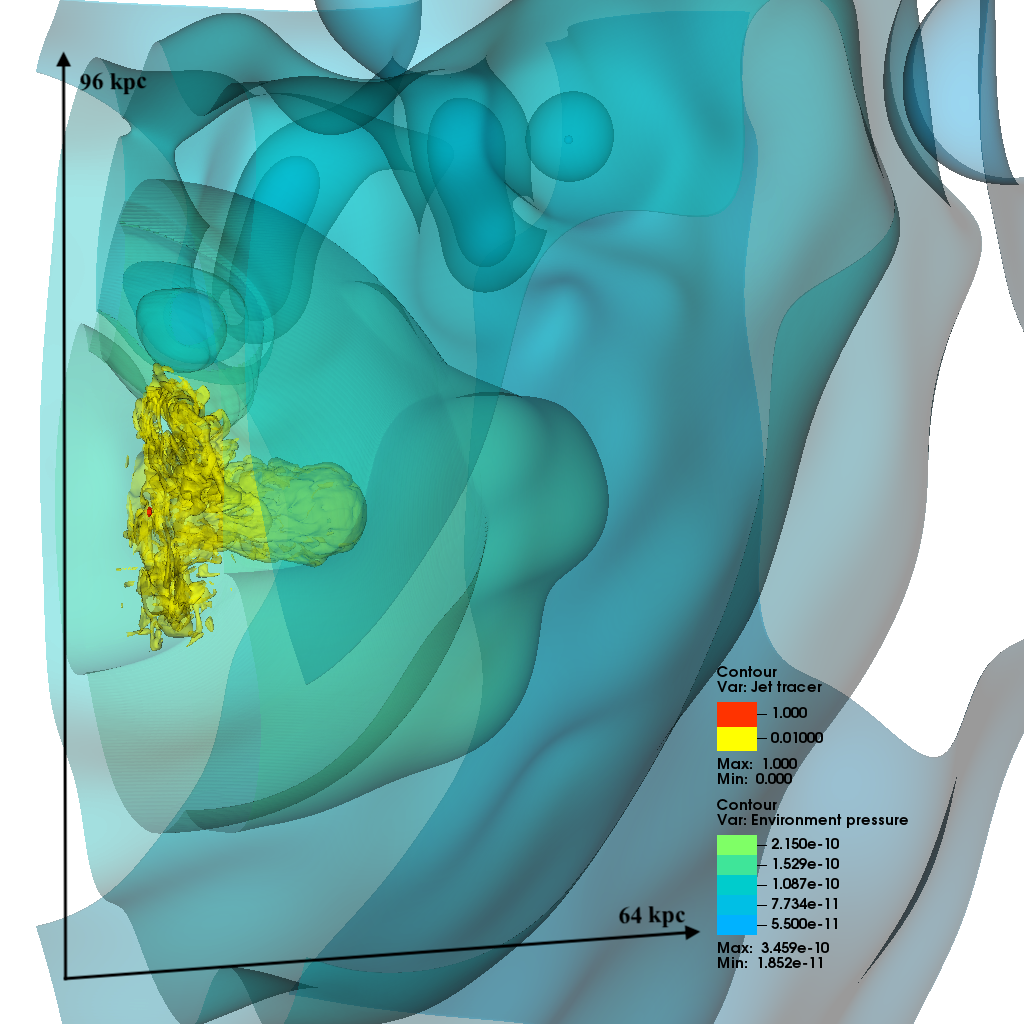}
    \includegraphics[width=0.44\linewidth]{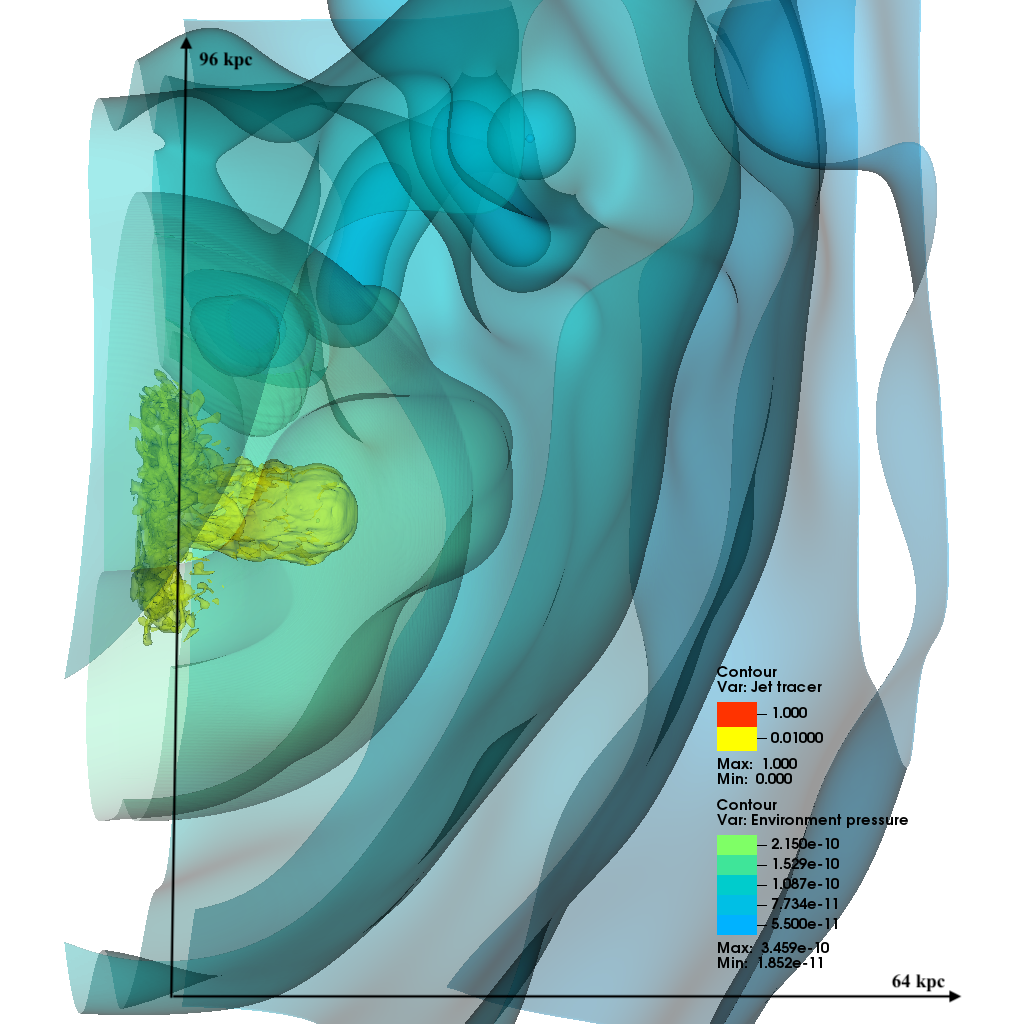}
    \caption{3D-RHD simulation of a 5$\times10^{44}$ erg/s relativistic jet propagating (from left to right) through an inhomogeneous galactic-to-cluster ambient medium, derived from cosmological simulations \citep{Planelles2014}. Jet mass fraction contours (yellow-red scale) track the propagation and mixing of injected material. Pressure contours (blue-green scale) for the environment gas included to show both the environment substructures and the position of the jet driven bow shock. The simulations have been run with the RHD code Ratpenat \citep[][Hervella-Seone et al., in preparation]{2010A&A...519A..41P}.}
    \label{fig:rhd-simulation}
\end{figure}
\section{Simulations and Preparatory Work }

Our current understanding of jet physics has been partially built on the contribution of numerical simulations and analytical models of jet evolution. 
Recent relativistic magnetohydrodynamic (RMHD) simulations have studied the evolution of jets through galactic atmospheres obtained from cosmological simulations \citep[e.g.][Hervella-Seoane et al., in prep.]{2023PASA...40...14Y}. This allows us to study jets with different compositions and properties (e.g. power, speed, opening angles) evolving through different environments (i.e. host galaxies at different redshift). Through these simulations (e.g. Fig.~\ref{fig:rhd-simulation}), we are able to study the evolution of jets and the physical processes dominating the interaction with the ambient medium, as well as to make predictions on the cosmological evolution of these processes. Exploring the parameter space of these simulations will allow us to better analyze the data gathered from active galaxies observed by the SKA, and this data can, in turn, be used to validate and improve on our models.

Another plausible and ongoing approach to this problem is the use of semi-analytical models to generate mock samples of jets, according to different power distributions, and also considering possible cosmological evolution \citep[][Beltran-Palau et al., in prep.]{2025arXiv250924939B}. The different parameters inferred from the simulated samples (e.g., linear size or radio luminosity) can be compared to observational distributions and use these fits as a test of the hypothesis used to generate them \citep[e.g.,][]{2015ApJ...806...59T,2018MNRAS.475.2768H,2025arXiv250924939B}
    
Using both devoted RMHD numerical simulations and simulated samples we can produce synthetic SKA observations of the simulated jets or make predictions on AGN jet detections by SKA based on different hypotheses on cosmological evolution.
The development of virtual twins to the SKA as part of SKAO developments, along with efforts by collaborations such as ODISSEE, will allow SKA users to generate synthetic SKA data faster than the simulated observing time. This will significantly strengthen the comparisons between simulation and observation by providing Observatory-supported tools to directly confront theoretical models with observational conditions and real observations.

\section{Observational Strategies and Early Science}

The synergy of the SKA and SKA-VLBI with current and future radio facilities, such as the ngVLA and ngEHT, will be crucial to provide sensitive, high-resolution maps of AGN jets from the collimation to the dissipation regions, and filling the gap in between current VLBI and connected-interferometer scales.
This goal can already be targeted in the early SKA Science phase with the AA* telescope for a pilot sample of well-studied and bright AGN. 
In particular, the combination of SKA-Mid with the international LOFAR telescope and eMERLIN will allow for detailed spectral studies from 150 MHz up to 10 GHz with similar angular resolutions of about 0.3 arcseconds, providing relevant information on the history of the AGN. At the same time, the SKA-VLBI will probe scales down to mas with high sensitivity, providing access to a new population of the faint AGN and their jets, whose emission is invisible to current VLBI instruments. 

In order to probe the faint AGN population, existing studies of the known low luminosity population can be extended through the SKA. The Palomar spectroscopic survey of 486 Northern galaxies \citep{Ho97}, complemented by existing radio detections from studies such as \cite{Nag05}, provides a suitable starting point for a faint AGN survey. SKA-Mid, in combination with SKA-Low and LOFAR will allow us to study radio signatures of the jets in LLAGN from MHz to GHz frequencies. Additionally, targeting fully multiwavelength fields such as the COSMOS field \citep{Scoville2007} is crucial in studying not only the radio properties of the faint radio-source populations and linking jets at small and large scales, but very importantly having the necessary multiwavelength observations to study the AGN jets in relation to small and large scale environments.

Observations combining the European VLBI Network (EVN) with e-MERLIN telescopes have already proved the mostly hidden capabilities of VLBI arrays which combine long-baselines with short-baselines, using the whole primary fov of the array elements to recover multiple structures within one observation. These capabilities use methods such as multiple phase center correlation and multi-scale imaging approaches. Adding SKA as a high-sensitivity long baseline to VLBI arrays, while obtaining standard SKA data products, will provide exactly this setup required to allow for a more detailed, higher-sensitivity study of larger FoVs. This is highly relevant for the expected high density of radio sources within the fov expected through SKAs high sensitivity. The standard SKA products, delivered with the VLBI observation, will allow us to pinpoint the brightest sources within the SKA primary fov. Through exploration of multiple phase-center correlation approaches for the SKA-VLBI observation, we have potential access to high-resolution images to several of these sources, allowing for a systematic study of galaxy groups and clusters containing AGN jets impacting the ambient medium through feedback.

\subsection{Early Science with AA*}

\begin{figure}[!ht]
  \begin{center}
    \includegraphics[width=\textwidth]{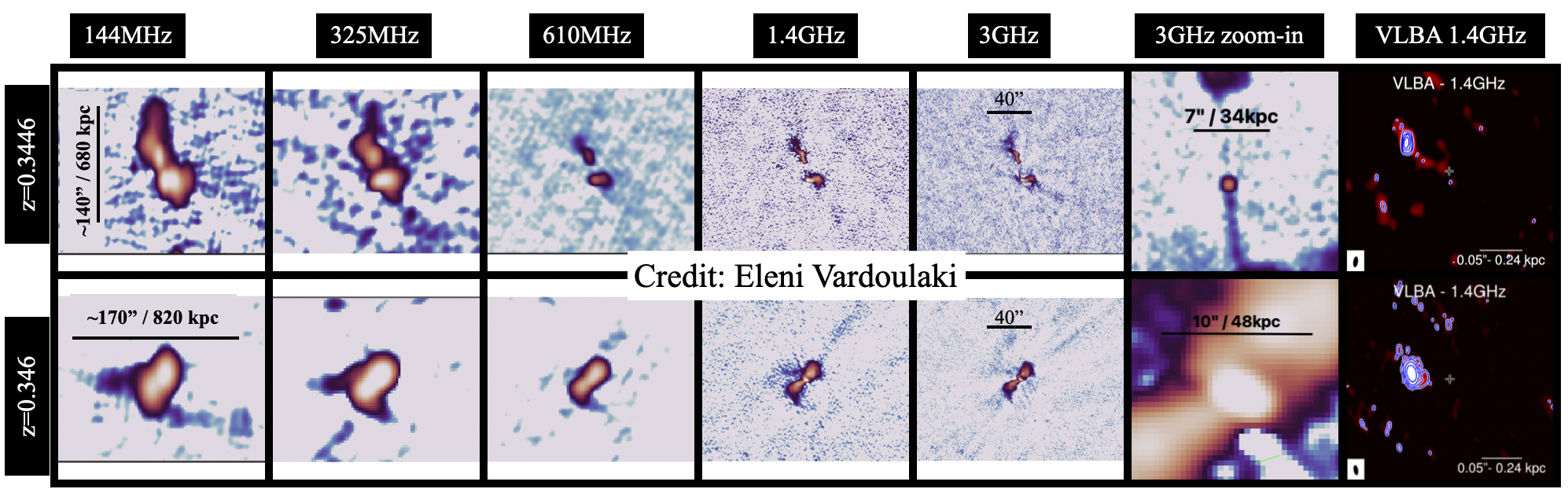}
  \end{center}
  \caption{\textit{Examples of two radio AGN from the COSMOS field  \citep[][ and Vardoulaki in prep.]{Vardoulaki2019,Vardoulaki2021a} as targets for early science with SKA-Mid AA*, ranging frequencies from 144 MHz to 3 GHz, including 1.4~GHz VLBA observations \citep{HerreraRuiz2017}}}
  \vspace{-0.5cm}
  \label{fig:earlyscience_COSMOS}
\end{figure}

Early science with SKA-Mid AA* would benefit from the inclusion of targets of opportunity in multiwavelength fields. We have selected two sources at redshifts $z \sim$ 0.34 which have multiwavelength coverage from low MHz with LOFAR to GHz frequencies with the VLA, including VLBA observations (see Fig.~\ref{fig:earlyscience_COSMOS}). These sources (a wide-angle-tail WAT and a X-shaped source) are members of X-ray group galaxies, they exhibit distortion in their radio structure, due to interaction with the environment, and they are studied from scales of $\sim$ 800 kpc down to 0.24 kpc. With the capabilities of SKA-Mid AA* we can resolve the jet at scales of 0.2" or $\sim$0.9~kpc, and bridge the VLBA and the LOFAR-VLBI (Vardoulaki et al. in prep.) and 3 GHz \citep{Vardoulaki2021a} shapes of these sources. 

\begin{figure}[!ht]
  \begin{center}
    \includegraphics[width=\textwidth]{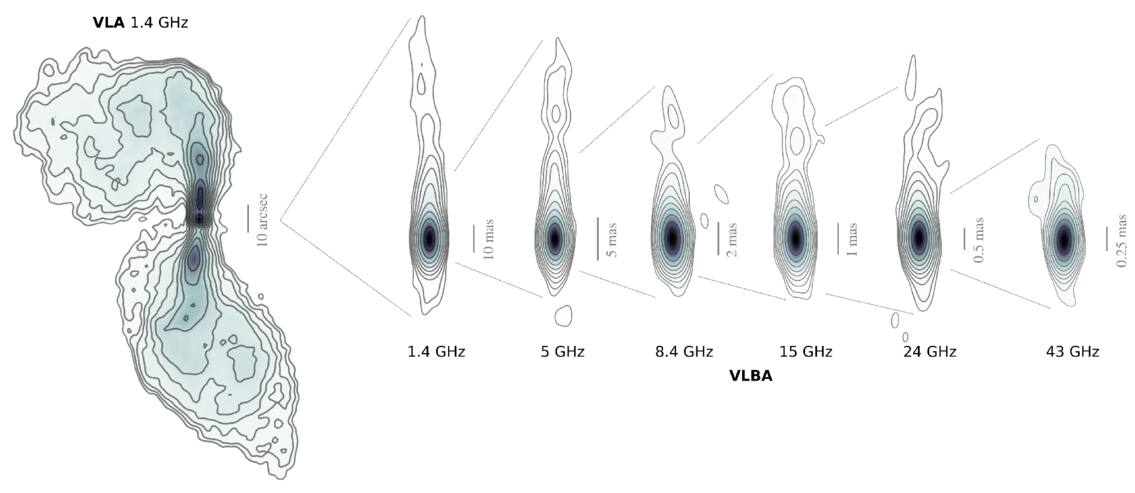}
  \end{center}
  \caption{\textit{Example of the low luminosity AGN M\,84 observed with the VLA and VLBA at multiple frequencies between 1.4 and 43\,GHz \citep{Far25} as target for early science with SKA-Mid AA*.}}
  \vspace{-0.5cm}
  \label{fig:earlyscience_M84}
\end{figure}

In order to allow for the study of jet expansion profiles from close to the formation region, covering the collimation, the break point to free expansion and the bondi radius, and connecting the inner jet to the larger scales jet expansion we selected one additional source whose combination of redshift and SMBH mass allows to resolve scales down to $10^3\,R_\mathrm{S}$. At a redshift of $z=0.0034$ M\,84 reveals large-scale structures in North-South direction in VLA observations, while the VLBA resolves the small scale jet at mas resolution \citep[See Fig. \ref{fig:earlyscience_M84}, taken from][]{Far25}. The combination of SKA-mid AA* as part of a global VLBI array will reach the desired sensitivity to also detect the extended emission with mas resolution, while the VLBA is limited in terms of high sensitivity, short baselines. The addition of SKA-Mid AA* will increase the North-South resolution by a factor of three, providing a circular beam. The significantly increased baseline sensitivities to SKA-mid AA* below $0.5\,\mathrm{mJy/beam}$ will guarantee detection on the long baselines and boost the image sensitivity. The coverage provided by e-MERLIN telescopes will further add short baselines to recover extended scales.

\section{Summary and Outlook}

The combination and inclusion of SKA-Low and SKA-Mid into global VLBI arrays will provide new insights into jet physics over all power-levels of AGN, from the bright to the faint population. By characterizing jet formation, propagation, and feedback mechanisms across spatial scales, we will inform cosmological models of AGN evolution - particularly in the low-luminosity regime, which dominates AGN activity over cosmic time. SKA will be transformative for the study of AGN jets in the following ways:
\begin{itemize}
    \item Continuous mapping of the radio structure of jets and distinction from winds from parsec to kiloparsec scales, relevant for jet propagation and collimation studies.
    \item Bridging the gap between compact and extended components, providing a complete view of how jets evolve and interact across physical scales
    \item Determine realistic source sizes and physical conditions ( e.g. FRI vs FRII dichotomy) across vastly larger populations of radio AGN
    \item Synergy with existing connected-interferometers will enable matched-resolution, multi-frequency studies, mapping electron populations from low-frequency fossil plasma to high-frequency emission regions
    \item Better understand of the process dominating the complex interaction between jets and the multiphase ISM and the IGM through SKAs resolution and sensitivity
    \item SKA will allow for statistical studies of so-far underrepresented AGN populations, which are too faint for current facilities or in the Southern hemisphere
    \item Upcoming SKA data will allow us to test, validate and improve RMHD models by giving a new view into pc-kpc scales.
\end{itemize}

Especially in combination with global VLBI arrays, SKA-Low and SKA-Mid will be groundbreaking for our understanding of the broader AGN population and the link between AGN jet activity and galaxy evolution.

\section*{Acknowledgements}

MP acknowledges support from the Astrophysics and High Energy Physics program supported by the Spanish Ministry of Science and Generalitat Valenciana with funding from European Union NextGenerationEU (\texttt{PRTR-C17.I1}) through grant \texttt{ASFAE/2022/005}, by the Spanish Ministry of Science through Grant \texttt{PID2022-136828NB-C43}, and by the Generalitat Valenciana through grant \texttt{CIPROM/2022/49};
J. Röder acknowledges financial support from the Severo Ochoa grant CEX2021-001131-S funded by MCIN/AEI/ 10.13039/501100011033;
LKM is grateful for support from a UKRI FLF [MR/Y020405/1];
JYK is supported for this research by the National Research Foundation of Korea (NRF) grant funded by the Korean government (Ministry of Science and ICT; grant no. 2022R1C1C1005255, RS-2022-NR071771) and by the Korea Astronomy and Space Science Institute under the R\&D program (Project No. 2025-9-844-00) supervised by the Korea AeroSpace Administration;
DGN acknowledges funding from Conicyt through Fondecyt Postdoctorado (project code 3220195), Comit\'e Mixto ESO-Chile, and N\'ucleo Milenio TITANs (projects NCN22{\_}002, NCN23{\_}002);
FU acknowledges support from the research project PRIN 2022 ``AGN-sCAN: zooming-in on the AGN-galaxy connection since the cosmic noon", contract 2022JZJBHM\_002 -- CUP J53D23001610006;
MBrienza acknowledges financial support from Next Generation EU funds within the National Recovery and Resilience Plan (PNRR), Mission 4 – Education and Research, Component 2 – From Research to Business (M4C2), Investment Line 3.1 – Strengthening and creation of Research Infrastructures, Project IR0000034 – “STILES – Strengthening the Italian Leadership in ELT and SKA”, from INAF under the Large GO 2024 funding scheme (project "MeerKAT and Euclid Team up: Exploring the galaxy-halo connection at cosmic noon”), the Large Grant 2022 funding scheme (project "MeerKAT and LOFAR Team up: a Unique Radio Window on Galaxy/AGN co-Evolution") and the Mini Grant 2023 funding scheme (project ‘Low radio frequencies as a probe of AGN jet feedback at low and high redshift’);
This research was funded by the Deutsche Forschungsgemeinschaft (DFG, German Research Foundation) as part of the DFG Research Unit FOR5195 – project number 443220636;
Jan Röder acknowledges financial support from the Severo Ochoa grant CEX2021-001131-S funded by MCIN/AEI/ 10.13039/501100011033;
The use of AI tools was limited to language editing assistance.

\bibliographystyle{abbrvnat-maxbibnames4}
\bibliography{chapter}

\end{document}

%% file: journal-names.tex
\newcommand{\actaa}{Acta Astron.} 
\newcommand{\araa}{ARA\&A} 
\newcommand{\aar}{A\&ARv} 
\newcommand{\aapr}{A\&ARv} 
\newcommand{\ab}{Astrobiol.} 
\newcommand{\aj}{AJ} 
\newcommand{\apj}{ApJ} 
\newcommand{\apjl}{ApJL} 
\newcommand{\apjs}{ApJSS} 
\newcommand{\ao}{Appl. Opt.} 
\newcommand{\apss}{Astro. \& Space Sci.} 
\newcommand{\aap}{A\&A} 
\newcommand{\aaps}{A\&AS.} 
\newcommand{\baas}{Bull. Am. Astron. Soc.} 
\newcommand{\caa}{Chinese A\&A} 
\newcommand{\cjaa}{Chinese J. A\&A} 
\newcommand{\cqg}{Class. Quantum Gravity} 
\newcommand{\gal}{Galaxies} 
\newcommand{\gca}{Geo. Cosmo. Acta} 
\newcommand{\icarus}{Icarus} 
\newcommand{\jcap}{JCAP} 
\newcommand{\jgr}{J. Geophys. Res.} 
\newcommand{\jgrp}{J. Geophys. Res. Planets} 
\newcommand{\jqsrt}{J. Quant. Spectrosc. Radiat. Transf.} 
\newcommand{\memsai}{Mem. SAIt} 
\newcommand{\mnras}{MNRAS} 
\newcommand{\nat}{Nature} 
\newcommand{\nastro}{Nat. Astron.} 
\newcommand{\ncomms}{Nat. Commun.} 
\newcommand{\nphys}{Nat. Phys.} 
\newcommand{\na}{New Astron.} 
\newcommand{\nar}{New Astron. Rev.} 
\newcommand{\physrep}{Phys. Rep.} 
\newcommand{\pra}{Phys. Rev. A} 
\newcommand{\prb}{Phys. Rev. B} 
\newcommand{\prc}{Phys. Rev. C} 
\newcommand{\prd}{Phys. Rev. D} 
\newcommand{\pre}{Phys. Rev. E} 
\newcommand{\prx}{Phys. Rev. X} 
\newcommand{\prl}{Phys. Rev. Let.} 
\newcommand{\psj}{Planet. Sci. J.} 
\newcommand{\planss}{Planet. Space Sci.} 
\newcommand{\pnas}{Proc. Natl Acad. Sci. USA} 
\newcommand{\procspie}{Proc. SPIE} 
\newcommand{\pasa}{PASA} 
\newcommand{\pasj}{PASJ} 
\newcommand{\pasp}{PASP} 
\newcommand{\rmxaa}{RMXAA} 
\newcommand{\sci}{Science} 
\newcommand{\sciadv}{Sci. Adv.} 
\newcommand{\solphys}{Sol. Phys.} 
\newcommand{\sovast}{Soviet Ast.} 
\newcommand{\ssr}{Space Sci. Rev.} 
\newcommand{\uni}{Universe} 